\documentclass[10pt,conference]{IEEEtran}
\IEEEoverridecommandlockouts

\usepackage{amsmath}
\usepackage{amssymb}
\usepackage{graphicx}
\usepackage{psfrag}
\usepackage{cite}
\usepackage{cases}

\graphicspath{{fig/}{num/}}
\newcommand{\E}{\mathrm{E}}
\newcommand{\EH}{\mathrm{E}_H}

\DeclareMathOperator{\dB}{dB}
\DeclareMathOperator{\CSIT}{CSIT}
\DeclareMathOperator{\rlz}{rlz}

\providecommand{\abs}[1]{\lvert{#1}\rvert}
\providecommand{\asq}[1]{\abs{#1}^2}

\begin{document}

\title{Minimum Expected Distortion in Gaussian Layered Broadcast Coding with Successive Refinement}

\author{%
  \authorblockN{Chris T. K. Ng\authorrefmark{1},
    Deniz G\"{u}nd\"{u}z\authorrefmark{2},
    Andrea J. Goldsmith\authorrefmark{1}, and
    Elza Erkip\authorrefmark{2}
  }\\
  \authorblockA{%
    \authorrefmark{1}Dept.\ of Electrical Engineering,
                     Stanford University, Stanford, CA 94305 USA 
  }
  \authorblockA{%
    \authorrefmark{2}Dept.\ of Electrical and Computer Engineering,
                     Polytechnic University, Brooklyn, NY 11201 USA
  }
  Email: \authorrefmark{1}\{ngctk,andrea\}@wsl.stanford.edu, 
  \authorrefmark{2}dgundu01@utopia.poly.edu, elza@poly.edu
  \thanks{This work was supported by the US Army under MURI award W911NF-05-1-0246, the ONR under award N00014-05-1-0168, DARPA under grant 1105741-1-TFIND, a grant from Intel, and the NSF under grant 0430885.}
}

\maketitle

%%% ============================================================
\begin{abstract}
%THIS PAPER IS ELIGIBLE FOR THE STUDENT PAPER AWARD.
A transmitter without channel state information (CSI) wishes to send a delay-limited Gaussian source over a slowly fading channel.
The source is coded in superimposed layers, with each layer successively refining the description in the previous one.
The receiver decodes the layers that are supported by the channel realization and reconstructs the source up to a distortion.
In the limit of a continuum of infinite layers, the optimal power distribution that minimizes the expected distortion is given by the solution to a set of linear differential equations in terms of the density of the fading distribution.
In the optimal power distribution, as SNR increases, the allocation over the higher layers remains unchanged; rather the extra power is allocated towards the lower layers.
On the other hand, as the bandwidth ratio $b$ (channel uses per source symbol) tends to zero, the power distribution that minimizes expected distortion converges to the power distribution that maximizes expected capacity.
While expected distortion can be improved by acquiring CSI at the transmitter (CSIT) or by increasing diversity from the realization of independent fading paths, at high SNR the performance benefit from diversity exceeds that from CSIT, especially when $b$ is large.
\end{abstract}

%%% ============================================================
\section{Introduction}

We consider the transmission of a delay-limited Gaussian source over a slowly fading channel in the absence of channel state information (CSI) at the transmitter.
As the channel is non-ergodic, source-channel separation is not necessarily optimal.
We consider the layered broadcast coding scheme in which each superimposed source layer successively refines the description in the previous one.
The receiver decodes the layers that are supported by the channel realization and reconstructs the source up to a distortion. 
We are interested in minimizing the expected distortion of the reconstructed source by optimally allocating the transmit power among the layers of codewords.

% 2) Related work
The broadcast strategy is proposed in \cite{cover72:broadcast_ch} to characterize the set of achievable rates when the channel state is unknown at the transmitter. 
In the case of a Gaussian channel under Rayleigh fading, \cite{shamai97:bc_strat_slow_fade, shamai03:bc_app_slow_fade_mimo} describe the layered broadcast coding approach and derive the optimal power allocation that maximizes the expected capacity.
% Hybrid
In the transmission of a Gaussian source over a Gaussian channel, uncoded transmission is optimal \cite{goblick65:lim_tx_analog_src} in the special case when the source bandwidth equals the channel bandwidth \cite{gastpar03:to_code_or_not}.
For other bandwidth ratios, hybrid digital-analog joint source-channel transmission schemes are studied in \cite{shamai98:sys_lossy_src_ch, mittal02:hda_src_ch_bc_rc, reznic06:dstrn_bnd_bc_bw_exp}, where the codes are designed to be optimal at a target SNR but degrade gracefully should the realized SNR deviate from the target.
%In particular, \cite{mittal02:hda_src_ch_bc_rc} conjectures that no code is simultaneously optimal at different SNRs when the source and channel bandwidths are not equal.
%In this paper, the code considered is not targeted for a specific fading state; we minimize the expected distortion over the fading distribution of the channel.

The distortion exponent, defined as the exponential decay rate of the expected distortion in the high SNR regime, is investigated in \cite{laneman05:src_ch_parl_ch} in the transmission of a source over two independently fading channels.
For quasi-static multiple-antenna Rayleigh fading channels, distortion exponent upper bounds and achievable joint source-channel schemes are studied in \cite{gunduz05:src_ch_code_quasi_fading, gunduz06:jt_src_ch_code_mimo, caire05:snr_expn_hybrid_st}.
The expected distortion of the layered source coding with progressive transmission (LS) scheme proposed in \cite{gunduz06:jt_src_ch_code_mimo} is analyzed in \cite{etemadi06:opt_layered_tx} for a finite number of layers at finite SNR\@.
Concatenation of broadcast channel coding with successive refinement \cite{equitz91:sus_refn_info, rimoldi94:sus_refn_info_ach} source coding is shown in \cite{gunduz05:src_ch_code_quasi_fading, gunduz06:jt_src_ch_code_mimo} to be optimal in terms of the distortion exponent for multiple input single output (MISO) and single input multiple output (SIMO) channels.
%The optimization of power and rate allocation among the layers is considered in \cite{zachariadis05:src_fid_fading}, and approximate solutions in the high SNR regime are presented.
Numerical optimization of the power allocation with constant rate among the layers is examined in \cite{sesia05:pro_sup_hyb}, while \cite{zachariadis05:src_fid_fading} considers
the optimization of power and rate allocation and presents approximate solutions in the high SNR regime.
The optimal power allocation that minimizes the expected distortion at finite SNR in layered broadcast coding is derived in \cite{ng07:recur_pow_lbc} when the channel has a finite number of discrete fading states. This work extends \cite{ng07:recur_pow_lbc} and considers the minimum expected distortion for channels with continuous fading distributions.
%Motivated by the optimality of the broadcast strategy in the high SNR regime, in this work we investigate the expected distortion at any arbitrary finite SNR\@.
In a related work in \cite{tian07:exp_dist_gaus_src_bc}, the optimal power distribution that minimizes the expected distortion is derived using the calculus of variations method.

The remainder of the paper is organized as follows.
Section~\ref{sec:sys_mod} presents the system model, 
and Section~\ref{sec:layered_bc_code} describes the layered broadcast coding scheme with successive refinement.
The optimal power distribution that minimizes the expected distortion is derived in
Section~\ref{sec:opt_pow_dist}.
Section~\ref{sec:ray_div} considers Rayleigh fading channels with diversity, followed by conclusions in Section~\ref{sec:conclu}.

%%% ============================================================
\section{System Model}
\label{sec:sys_mod}

\begin{figure}
  \centering
  \psfrag{r~f(r)}[][]{$\gamma \sim f(\gamma)$}
  \psfrag{xN}[][]{$x^N$}
  \psfrag{yN}[][]{$y^N$}
  \psfrag{sK}[][]{$s^K$}
  \psfrag{sKh}[][]{$\hat{s}^K$}
  \psfrag{CN(0,1)}[][]{$\mathcal{CN}(0,1)$}
  \includegraphics{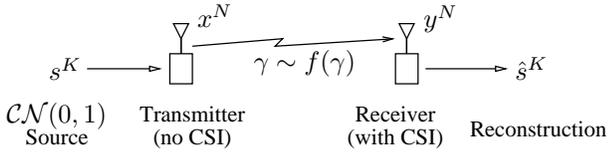}
  \caption{Source-channel coding without CSI at the transmitter.}
  \label{fig:src_ch_coding_pdf}
\end{figure}

Consider the system model illustrated in Fig.~\ref{fig:src_ch_coding_pdf}: A transmitter wishes to send a Gaussian source over a wireless channel to a receiver, at which the source is to be reconstructed with a distortion.
Let the source be denoted by $s$, which is a sequence of independent identically distributed (iid) zero-mean circularly symmetric complex Gaussian (ZMCSCG) random variables with unit variance: $s\in\mathbb{C}\sim\mathcal{CN}(0,1)$.
The transmitter and the receiver each have a single antenna and the channel is described by:
%\begin{align}
  $y = Hx + n$,
%\end{align}
where $x\in\mathbb{C}$ is the transmit signal, $y\in\mathbb{C}$ is the received signal, and $n\in\mathbb{C}\sim\mathcal{CN}(0,1)$ is iid unit-variance ZMCSCG noise.

Suppose the distribution of the channel power gain is described by the probability density function (pdf) $f(\gamma)$, where $\gamma \triangleq \asq{h}$ and $h\in\mathbb{C}$ is a realization of $H$.
The receiver has perfect CSI but the transmitter has only channel distribution information (CDI), i.e., the transmitter knows the pdf $f(\gamma)$ but not its instantaneous realization.
The channel is modeled by a quasi-static block fading process: $H$ is realized iid at the onset of each fading block and remains unchanged over the block duration.
We assume decoding at the receiver is \emph{delay-limited};
namely, delay constraints preclude coding across fading blocks but dictate that the receiver decodes at the end of each block.
Hence the channel is non-ergodic. 

Suppose each fading block spans $N$ channel uses, over which the transmitter describes $K$ of the source symbols. We define the \emph{bandwidth ratio} as $b\triangleq N/K$, which relates the number of channel uses per source symbol.
At the transmitter there is a power constraint on the transmit signal $\E\bigl[\asq{x}\bigr] \leq P$, where the expectation is taken over repeated channel uses over the duration of each fading block.
We assume a short-term power constraint and do not consider power allocation across fading blocks.
We assume $K$ is large enough to consider the source as ergodic, and $N$ is large enough to design codes that achieve the instantaneous channel capacity of a given fading state with negligible probability of error.

At the receiver, the channel output $y$ is used to reconstruct an estimate $\hat{s}$ of the source.
The distortion $D$ is measured by the mean squared error $\E[(s-\hat{s})^2]$ of the estimator, where the expectation is taken over the $K$-sequence of source symbols and the noise distribution.
The instantaneous distortion of the reconstruction depends on the fading realization of the channel; we are interested in minimizing the expected distortion $\EH[D]$, where the expectation is over the fading distribution.

%%% ============================================================
\section{Layered Broadcast Coding with\\Successive Refinement}
\label{sec:layered_bc_code}

\begin{figure}
  \centering
  \psfrag{p1:r1}[][]{$p_1:\gamma_1$}
  \psfrag{p2:r2}[][]{$p_2:\gamma_2$}
  \psfrag{pM:rM}[][]{$p_M:\gamma_M$}
  \psfrag{(P1,R1)}[][]{$(P_1,R_1)$}
  \psfrag{(P2,R2)}[][]{$(P_2,R_2)$}
  \psfrag{(PM,RM)}[][]{$(P_M,R_M)$}
  \psfrag{sK}[][]{$s^K$}
  \psfrag{sKh}[][]{$\hat{s}^K$}
  \includegraphics{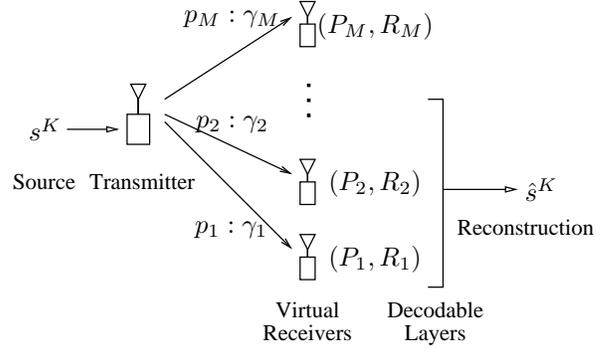}
  \caption{Layered broadcast coding with successive refinement.}
  \label{fig:src_ch_layers}
\end{figure}

We build upon the power allocation framework derived in \cite{ng07:recur_pow_lbc}, and first assume the fading distribution has $M$ discrete states: the channel power gain realization is $\gamma_i$ with probability $p_i$, for $i=1,\dotsc,M$, as depicted in Fig.~\ref{fig:src_ch_layers}. 
Accordingly there are $M$ virtual receivers and the transmitter sends the sum of $M$ layers of codewords.
Let layer~$i$ denote the layer of codeword intended for virtual receiver~$i$,
and we order the layers as $\gamma_M>\dotsb>\gamma_1\geq0$.
We refer to layer~$M$ as the highest layer and layer~1 as the lowest layer.
Each layer successively refines the description of the source $s$ from the layer below it, and the codewords in different layers are independent.
Let $P_i$ be the transmit power allocated to layer~$i$, then the transmit symbol $x$ can be written as
\begin{align}
x &= \sqrt{P_1}\,x_1 + \sqrt{P_2}\,x_2 + \dotsb +\sqrt{P_M}\,x_M,
\end{align}
where $x_1,\dotsc,x_M$ are iid ZMCSCG random variables with unit variance. 
Suppose the layers are evenly spaced, with $\gamma_{i+1}-\gamma_i = \Delta\gamma$.
In Section~\ref{sec:opt_pow_dist} we consider the limiting process as $\Delta\gamma\rightarrow0$ to obtain the power distribution:
\begin{align}
\rho(\gamma) \triangleq \lim_{\Delta\gamma \rightarrow 0} 
\dfrac{1}{\Delta\gamma} P_{\lceil \gamma/\Delta\gamma \rceil},
\end{align}
where for discrete layers the power allocation $P_i$ is referenced by the integer layer index $i$, while the continuous power distribution $\rho(\gamma)$ is indexed by the channel power gain $\gamma$.

With successive decoding \cite{cover91:eoit}, each virtual receiver first decodes and cancels the lower layers before decoding its own layer; the undecodable higher layers are treated as noise. Thus the rate $R_i$ intended for virtual receiver~$i$ is
\begin{align}
R_i &= \log\biggl(1+\frac{\gamma_i P_i}{1+\gamma_i \sum_{j=i+1}^{M}P_j}\biggr),
\end{align}
where the term $\gamma_i \sum_{j=i+1}^{M}P_j$ represents the interference power from the higher layers.
Suppose $\gamma_k$ is the realized channel power gain, then the original receiver can decode layer~$k$ and all the layers below it. Hence the realized rate $R_{\rlz}(k)$ at the original receiver is $R_1+\dotsb+R_k$.

From the rate distortion function of a complex Gaussian source \cite{cover91:eoit}, the mean squared distortion is $2^{-bR}$ when the source is described at a rate of $bR$ per symbol.
Thus the realized distortion $D_{\rlz}(k)$ of the reconstructed source $\hat{s}$ is
\begin{align}
  D_{\rlz}(k) &= 2^{-bR_{\rlz}(k)} 
        = 2^{-b(R_1+\dotsb+R_k)},
\end{align}
where the last equality follows from successive refinability \cite{equitz91:sus_refn_info, rimoldi94:sus_refn_info_ach}.
The expected distortion $\EH[D]$ is obtained by averaging over the fading distribution:
\begin{align}
%\label{eq:exp_dist_pR}
\EH[D] &= \sum_{i=1}^{M} p_i D_{\rlz}(i)
%= \sum_{i=1}^M p_i 2^{-b(\sum_{j=1}^i R_j)}\\
\label{eq:ED_sum_prods}
= \sum_{i=1}^M p_i \Bigl(\prod_{j=1}^i \frac{1+\gamma_j T_j}{1+\gamma_j T_{j+1}}\Bigr)^{-b},
\end{align}
where $T_i$ represents the cumulative power in layers $i$ and above:
%\begin{align}
%T_i \triangleq \sum_{j=i}^{M}P_j,\;\text{for $i=1,\dotsc,M$};
%\quad T_{M+1}\triangleq 0.
%\end{align}
$T_i \triangleq \sum_{j=i}^{M}P_j$, for $i=1,\dotsc,M$; $T_{M+1}\triangleq 0$.
In the next section we derive the optimal cumulative power allocation $T_2^*,\dotsc,T_M^*$ to find the minimum expected distortion $\EH[D]^*$.
%Note that the expected distortion is monotonically decreasing in the transmit power $P$, hence the power constraint can be taken as an equality $\sum_{i=1}^{M} P_i = P$, and the optimization formulated as:
%%\begin{align}
%%\begin{split}
%%\EH[D]^* &= 
%%\min_{P_1,\dotsc,P_M}\EH[D]\\
%%&\text{subject to } P_i\geq 0,\,{\textstyle\sum P_i}= P,\;\forall i=1,\dotsc,M.
%%\end{split}
%%\end{align}
%\begin{align}
%\begin{split}
%\EH[D]^* &= 
%\min_{T_2,\dotsc,T_M}\EH[D]\\
%&\text{subject to } 0\leq T_M\leq\dotsb\leq T_2\leq P.
%\end{split}
%\end{align}

%Power constraint becomes:
%\begin{align}
%\int_0^\infty \rho(\gamma) \,d\gamma \leq P
%\end{align}

%%% ============================================================
\section{Optimal Power Distribution}
\label{sec:opt_pow_dist}

To derive the minimum expected distortion, we factor the sum of cumulative products in (\ref{eq:ED_sum_prods}) and rewrite the expression as a set of recurrence relations:
%\begin{align}
%D_M &\triangleq \bigl(1+\gamma_M T_M\bigr)^{-b}p_M\\
%D_i &= \Bigl(\frac{1+\gamma_i T_i}{1+\gamma_i T_{i+1}}\Bigr)^{-b}\bigl(p_i+D_{i+1}\bigr),
%\end{align}
\begin{align}
D_M^* &\triangleq \bigl(1+\gamma_M T_M\bigr)^{-b}p_M\\
\label{eq:recur_min_dist}
D_i^* 
&=\min_{0\leq T_{i+1} \leq T_i}
\Bigl(\frac{1+\gamma_i T_i}{1+\gamma_i T_{i+1}}\Bigr)^{-b}\bigl(p_i+D_{i+1}^*\bigr),
\end{align}
where $i$ runs from $M-1$ down to 1. The term $D_i^*$ can be interpreted as the cumulative distortion from layers~$i$ and above, with $D_1^*$ equal to the minimum expected distortion $\EH[D]^*$.
Note that $D_i$ depends on only two adjacent power allocation variables $T_i$ and $T_{i+1}$; therefore, in each recurrence step $i$ in (\ref{eq:recur_min_dist}), we solve for the optimal $T_{i+1}^*$ in terms of $T_i$.

\begin{figure}
  \centering
  \psfrag{W(r):r}[][]{$W(\gamma):\gamma$}
  \psfrag{f(r)dr:r-dr}[][]{$f(\gamma)\Delta\gamma:\gamma-\Delta\gamma$}
  \psfrag{-T(r)}[][]{$-\,T(\gamma)$}
  \psfrag{T(r)}[][]{$T(\gamma)$}
  \psfrag{T(r-dr)}[][]{$T(\gamma-\Delta\gamma)$}
  \includegraphics{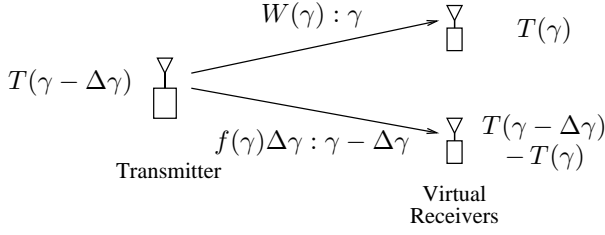}
  \caption{Power allocation between two adjacent layers.}
  \label{fig:two_delta_layers}
\end{figure}

Specifically, consider the optimal power allocation between layer~$\gamma$ and its lower layer $\gamma-\Delta\gamma$ as shown in Fig.~\ref{fig:two_delta_layers}.
Let $T(\gamma-\Delta\gamma)$ denote the available transmit power for layers~$\gamma-\Delta\gamma$ and above, of which $T(\gamma)$ is allocated to layers~$\gamma$ and above; the remaining power $T(\gamma)-T(\gamma-\Delta\gamma)$ is allocated to layer~$\gamma-\Delta\gamma$.
Under optimal power allocation, it is shown in \cite{ng07:recur_pow_lbc} that the cumulative distortion from layers $\gamma$ and above can be written in the form:
\begin{align}
\label{eq:Dr_W_form}
D^*(\gamma) = \bigl(1+\gamma T(\gamma)\bigr)^{-b} W(\gamma),
\end{align}
where $W(\gamma)$ is interpreted as an equivalent probability weight summarizing the aggregate effect of the layers~$\gamma$ and above.
For the lower layer in Fig.~\ref{fig:two_delta_layers}, $f(\gamma)\Delta\gamma$ represents the probability that layer~$\gamma-\Delta\gamma$ is realized.

In the next recurrence step as prescribed by (\ref{eq:recur_min_dist}), the cumulative distortion for the lower layer is
\begin{align}
&D^*(\gamma-\Delta\gamma) = \min_{0\leq T(\gamma)\leq T(\gamma-\Delta\gamma)} D(\gamma-\Delta\gamma)\\
\begin{split}
\label{eq:Drd_min_Tr}
& \quad= \min_{0\leq T(\gamma)\leq T(\gamma-\Delta\gamma)}\Bigl(\frac{1+(\gamma-\Delta\gamma) T(\gamma-\Delta\gamma)}{1+(\gamma-\Delta\gamma) T(\gamma)}\Bigr)^{-b}\\
 &\hspace{7.25em}\cdot\Bigl[f(\gamma)\Delta\gamma+\bigl(1+\gamma T(\gamma)\bigr)^{-b} W(\gamma)\Bigr].
\end{split}
\end{align}
%The minimization can be solved by the Lagrange method. We form the Lagrangian:
We solve the minimization by forming the Lagrangian:
\begin{align}
\begin{split}
&L(T(\gamma),\lambda_1,\lambda_2) = \\
&\quad D(\gamma-\Delta\gamma) + \lambda_1\bigl(T(\gamma)-T(\gamma-\Delta\gamma)\bigr) - \lambda_2 T(\gamma).
\end{split}
\end{align}
The Karush-Kuhn-Tucker (KKT) conditions stipulate that the gradient of the Lagrangian vanishes at the optimal power allocation $T^*(\gamma)$, which leads to the solution:
%\begin{align}
%T^*(\gamma) = \min\bigl(U(\gamma),T(\gamma-\Delta\gamma)\bigr)
%\end{align}
\begin{subnumcases}
  {\label{eq:Tr_opt} T^*(\gamma) = }
  \label{eq:Tr_opt_uncon}
  U(\gamma) & if $U(\gamma) \leq T(\gamma-\Delta\gamma)$\IEEEeqnarraynumspace\\
  \label{eq:Tr_opt_con}
  T(\gamma-\Delta\gamma) & else,
\end{subnumcases}
where
\begin{subnumcases}
{\hspace{1em}U(\gamma) \triangleq}
    \label{eq:Ur_o}
    0 \text{\hspace{5.25em}if $\gamma \geq W(\gamma)/f(\gamma) + \Delta\gamma$} & \\
    \label{eq:Ur_nz}
    \dfrac{1}{\gamma}\biggl(\Bigl[\frac{W(\gamma)}{f(\gamma)(\gamma-\Delta\gamma)}\Bigr]^{\frac{1}{1+b}}-1\biggr)  \text{\hspace{2em}else.}&\hspace{1em}
\end{subnumcases}

We assume there is a region of $\gamma$ where the cumulative power allocation is not constrained by the power available from the lower layers, i.e., $U(\gamma)\leq U(\gamma-\Delta\gamma)$ and $U(\gamma)\leq P$.
In this region the optimal power allocation $T^*(\gamma)$ is given by the unconstrained minimizer $U(\gamma)$ in (\ref{eq:Tr_opt_uncon}).
In the solution to $U(\gamma)$ we need to verify that $U(\gamma)$ is non-increasing in this region, which corresponds to the power distribution $\rho^*(\gamma)$ being non-negative.
With the substitution of the unconstrained cumulative power allocation $U(\gamma)$ in (\ref{eq:Drd_min_Tr}), the cumulative distortion at layer $\gamma-\Delta\gamma$ becomes:
\begin{align}
\begin{split}
\label{eq:Drd_U_W}
D^*(\gamma-\Delta\gamma) &=
\Bigl(\frac{1+(\gamma-\Delta\gamma) T(\gamma-\Delta\gamma)}{1+(\gamma-\Delta\gamma) U(\gamma)}\Bigr)^{-b}\\
&\hspace{1.75em}\cdot\Bigl[f(\gamma)\Delta\gamma+\bigl(1+\gamma U(\gamma)\bigr)^{-b} W(\gamma)\Bigr],
\end{split}
\end{align}
which is of the form in (\ref{eq:Dr_W_form}) if we define $W(\gamma-\Delta\gamma)$ by the recurrence equation:
\begin{align}
\begin{split}
\label{eq:Wrd_Wr}
W(\gamma-\Delta\gamma) &= 
\bigl(1+(\gamma-\Delta\gamma) U(\gamma)\bigr)^b\\
&\hspace{2em}\cdot\bigl[f(\gamma)\Delta\gamma+\bigl(1+\gamma U(\gamma)\bigr)^{-b} W(\gamma)\bigr].
\end{split}
\end{align}

Next we consider the limiting process as the spacing between the layers condenses.
In the limit of $\Delta\gamma$ approaching zero, the recurrence equations (\ref{eq:Drd_U_W}), (\ref{eq:Wrd_Wr}) become differential equations.
The optimal power distribution $\rho^*(\gamma)$ is given by the derivative of the cumulative power allocation:
\begin{align}
\rho^*(\gamma) &= -{T^*}'(\gamma),
\end{align}
where $T^*(\gamma)$ is described by solutions in three regions:
\begin{subnumcases}
{\label{eq:Tr_all_opt}T^*(\gamma) =}
\label{eq:Tr_ro}
    0 & $\gamma > \gamma_o$ \\
\label{eq:Tr_rP_ro}
    U(\gamma) & $\gamma_P \leq \gamma \leq \gamma_o$\\
\label{eq:Tr_rP}
    P & $\gamma < \gamma_P$.
\end{subnumcases}
In region (\ref{eq:Tr_ro}) when $\gamma > \gamma_o$, corresponding to cases (\ref{eq:Tr_opt_uncon}) and (\ref{eq:Ur_o}), no power is allocated to the layers and (\ref{eq:Wrd_Wr}) simplifies to $W(\gamma) = 1-F(\gamma)$, where $F(\gamma) \triangleq \int_0^\gamma f(s)\,ds$ is the cumulative distribution function (cdf) of the channel power gain.
The boundary $\gamma_0$ is defined by the condition in (\ref{eq:Ur_o}) which satisfies:
\begin{align}
\label{eq:ro_rf_F}
\gamma_o f(\gamma_o) + F(\gamma_o) - 1 = 0.
\end{align}
Under Rayleigh fading when $f(\gamma) = \bar{\gamma}^{-1}e^{-\gamma/\bar{\gamma}}$, where $\bar{\gamma}$ is the expected channel power gain, (\ref{eq:ro_rf_F}) evaluates to $\gamma_o = \bar{\gamma}$.
For other fading distributions, $\gamma_o$ may be computed numerically.

In region (\ref{eq:Tr_rP_ro}) when $\gamma_P \leq \gamma \leq \gamma_o$, corresponding to cases (\ref{eq:Tr_opt_uncon}) and (\ref{eq:Ur_nz}), the optimal power distribution is described by a set of differential equations.
We apply the first order binomial expansion $(1+\Delta\gamma)^b\cong1+b\Delta\gamma$, and (\ref{eq:Wrd_Wr}) becomes:
\begin{align}
W'(\gamma) &= \lim_{\Delta\gamma \rightarrow 0} \frac{W(\gamma) - W(\gamma-\Delta\gamma)}{\Delta\gamma}\\
\label{eq:dW_W}
&= b\frac{W(\gamma)}{\gamma} - (1+b)\Bigl[f(\gamma)\Big(\frac{W(\gamma)}{\gamma}\Bigr)^b\Bigr]^{\frac{1}{1+b}},
\end{align}
which we substitute in (\ref{eq:Ur_nz}) to obtain:
\begin{align}
\label{eq:dU_U}
U'(\gamma) &= -\Big(\frac{2/\gamma + f'(\gamma)/f(\gamma)}{1+b}\Bigr) \Big[U(\gamma)+1/\gamma\Bigr].
\end{align}
Hence $U(\gamma)$ is described by a first order linear differential equation.
With the initial condition $U(\gamma_o) = 0$, its solution is given by
\begin{align}
\label{eq:Ur_int_fr}
U(\gamma) &= \frac{\displaystyle -\int_{\gamma_o}^{\gamma} 
\dfrac{1}{s}\Bigl(\dfrac{2}{s}+\dfrac{f'(s)}{f(s)}\Bigr) \bigl[s^2 f(s)\bigr]^{\frac{1}{1+b}} \,ds}
{(1+b)\bigl[\gamma^2 f(\gamma)\bigr]^{\frac{1}{1+b}}},
\end{align}
and condition (\ref{eq:Tr_opt_con}) in the lowest active layer becomes the boundary condition $U(\gamma_P) = P$.
% Added
In \cite{tian07:exp_dist_gaus_src_bc}, the power distribution in (\ref{eq:Ur_int_fr}) is derived using the calculus of variations method.

Similarly, as $\Delta\gamma\rightarrow0$, the evolution of the expected distortion in (\ref{eq:Drd_U_W}) becomes:
\begin{align}
D'(\gamma) &= -\dfrac{b\gamma U'(\gamma)}{1+\gamma U(\gamma)}D(\gamma) - f(\gamma)\\
 &= \Bigl[\dfrac{b}{1+b}\Bigl(\dfrac{2}{\gamma}+\dfrac{f'(\gamma)}{f(\gamma)}\Bigr)\Bigr]D(\gamma)- f(\gamma),
\end{align}
%Sub in $U'$ in second step.
which is again a first order linear differential equation.
With the initial condition $D(\gamma_o) = W(\gamma_o) = \gamma_o f(\gamma_o)$, its solution is given by
\begin{align}
D(\gamma) &= \frac{\displaystyle -\int_{\gamma_o}^{\gamma} f(s) 
\Bigl[\Bigl(\dfrac{s}{\gamma_o}\Bigr)^2\dfrac{f(s)}{f(\gamma_o)}\Bigr]^{\frac{-b}{1+b}} \,ds 
+ \gamma_o f(\gamma_o)}
{\Bigl[\Bigl(\dfrac{\gamma}{\gamma_o}\Bigr)^2\dfrac{f(\gamma)}{f(\gamma_o)}\Bigr]^{\frac{-b}{1+b}}}.
\end{align}

%%%%% injournal: Give these in the journal version
%For example, under Rayleigh fading, 
%%$f(\gamma)=\bar{\gamma}^{-1}e^{-\gamma/\bar{\gamma}}$, $\gamma_o = \bar{\gamma}$,
%\begin{align}
%\label{eq:U_ray}
%U(\gamma) &= \frac{\displaystyle \int_{\bar{\gamma}}^{\gamma}
%\Bigl(\dfrac{1}{\bar{\gamma}}-\dfrac{2}{s}\Bigr) \bigl[s^{1-b}e^{-s/\bar{\gamma}}\bigr]^{\frac{1}{1+b}} \,ds
%}
%{(1+b)\bigl[\gamma^2 e^{-\gamma/\bar{\gamma}}\bigr]^{\frac{1}{1+b}}},
%\end{align}
%\begin{align}
%\label{eq:D_ray}
%D(\gamma) &= \frac{\displaystyle -\dfrac{1}{\bar{\gamma}}\int_{\bar{\gamma}}^{\gamma} e^{-s/\bar{\gamma}} 
%\Bigl[\Bigl(\dfrac{s}{\bar{\gamma}}\Bigr)^2 e^{-(s-\bar{\gamma})/\bar{\gamma}}\Bigr]^{\frac{-b}{1+b}} \,ds 
%+ e^{-1}}
%{\Bigl[\Bigl(\dfrac{\gamma}{\bar{\gamma}}\Bigr)^2 e^{-(\gamma-\bar{\gamma})/\bar{\gamma}}\Bigr]^{\frac{-b}{1+b}}},
%\end{align}
%and the integrals in (\ref{eq:U_ray}), (\ref{eq:D_ray}) can be computed by evaluating the incomplete gamma function.

Finally, in region (\ref{eq:Tr_rP}) when $\gamma < \gamma_P$, corresponding to case (\ref{eq:Tr_opt_con}), the transmit power $P$ has been exhausted, and no power is allocated to the remaining layers.
%In this region $D(\gamma) = \int_\gamma^{\gamma_P} f(s) \,ds + D(\gamma_P)$; thus given by:
Hence the minimum expected distortion is
\begin{align}
\EH[D]^* = D(0) = F(\gamma_P) + D(\gamma_P),
\end{align}
where the last equality follows from when $\gamma < \gamma_P$ in region (\ref{eq:Tr_rP}), $\rho^*(\gamma)=0$ and
$D(\gamma) = \int_\gamma^{\gamma_P} f(s) \,ds + D(\gamma_P)$.

%%% ============================================================
\section{Rayleigh Fading with Diversity}
\label{sec:ray_div}

In this section we consider the optimal power distribution and the minimum expected distortion when the wireless channel undergoes Rayleigh fading with a diversity order of $L$ from the realization of independent fading paths.
Specifically, we assume the fading channel is characterized by the Erlang distribution:
\begin{align}
f_L(\gamma) = \frac{(L/\bar{\gamma})^L \gamma^{L-1} e^{-L\gamma/\bar{\gamma}} }
{(L-1)!}, \qquad\gamma > 0,
\end{align}
which corresponds to the average of $L$ iid channel power gains, each under Rayleigh fading with an expected value of $\bar{\gamma}$.
The $L$-diversity system may be realized by having $L$ transmit antennas using isotropic inputs, by relaxing the decode delay constraint over $L$ fading blocks, or by having $L$ receive antennas under maximal-ratio combining when the power gain of each antenna is normalized by $1/L$.

%%% Journal version:
%; the upper and lower boundaries $\gamma_o, \gamma_P$ of the span of the active layers are plotted in Fig.~\ref{fig:erlang_r0_rP}.
%As the diversity $L$ increases, the power distribution becomes more concentrated, albeit slowly.
%By the law of large numbers, at asymptotically large $L$, we expect all power concentrates at $\bar{\gamma}$.
%\begin{figure}
%  \centering
%  \includegraphics*[width=8cm]{fig_erlang_r0_rP_bw.eps}
%  \caption{Span of active layers under optimal power distribution.}
%  \label{fig:erlang_r0_rP}
%\end{figure}

\begin{figure}
  \centering
  \includegraphics*[width=8cm]{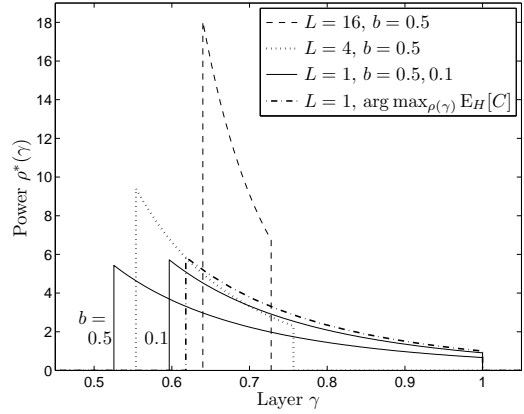}
  \caption{Optimal power distribution ($P=0~\dB$).}
  \label{fig:erlang_pow_dist}
\end{figure}

Fig.~\ref{fig:erlang_pow_dist} shows the optimal power distribution $\rho^*(\gamma)$,
%The optimal power distribution (\ref{eq:Tr_opt}) concentrates the transmit power over a range of active layers.
which is concentrated over a range of active layers.
A higher SNR $P$ or a larger bandwidth ratio $b$ extends the span of the active layers further into the lower layers but the upper boundary $\gamma_o$ remains unperturbed.
It can be observed that a smaller bandwidth ratio $b$ reduces the spread of the power distribution.
In fact, as $b$ approaches zero, the optimal power distribution that minimizes expected distortion converges to the power distribution that maximizes expected capacity.
To show the connection, we take the limit in the distortion-minimizing cumulative power distribution in (\ref{eq:Ur_int_fr}):
\begin{align}
\lim_{b\rightarrow0}U(\gamma) &= \frac{1-F(\gamma)-\gamma f(\gamma)}{\gamma^2 f(\gamma)},
\end{align}
which is equal to the capacity-maximizing cumulative power distribution as derived in \cite{shamai03:bc_app_slow_fade_mimo}.
Essentially, from the first order expansion $e^b\cong1+b$ for small $b$, $\EH[D]\cong 1- b \EH[C]$ when the bandwidth ratio is small, where $\EH[C]$ is the expected capacity in nats/s, and hence minimizing expected distortion becomes equivalent to maximizing expected capacity.
For comparison, the capacity-maximizing power distribution is also plotted in Fig.~\ref{fig:erlang_pow_dist}.
Note that the distortion-minimizing power distribution is more conservative, and it is more so as $b$ increases, as the allocation favors lower layers in contrast to the capacity-maximizing power distribution.

\begin{figure}
  \centering
  \includegraphics*[width=8cm]{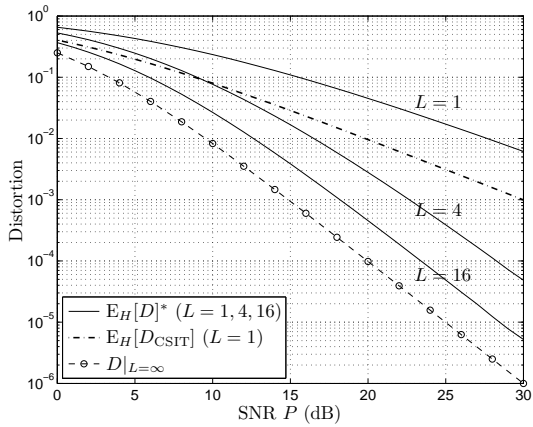}
  \caption{Minimum expected distortion ($b=2$).}
  \label{fig:erlang_ED_P}
\end{figure}

Fig.~\ref{fig:erlang_ED_P} shows the minimum expected distortion $\EH[D]^*$ versus SNR for different diversity orders.
With infinite diversity, the channel power gain becomes constant at $\bar{\gamma}$, and the distortion is given by
\begin{align}
D\vert_{L=\infty} = (1+\bar{\gamma}P)^{-b}.
\end{align}
In the case when there is no diversity ($L=1$), a lower bound to the expected distortion is also plotted.
The lower bound assumes the system has CSI at the transmitter (CSIT), 
which allows the transmitter to concentrate all power at the realized layer to achieve the expected distortion:
\begin{align}
\EH[D_{\CSIT}] &= \int_0^\infty e^{-\gamma}(1+\gamma P)^{-b}\,d\gamma.
\end{align}
Note that at high SNR, the performance benefit from diversity exceeds that from CSIT, especially when the bandwidth ratio $b$ is large.
In particular, in terms of the distortion exponent $\Delta$ \cite{laneman05:src_ch_parl_ch},
it is shown in \cite{gunduz06:jt_src_ch_code_mimo} that in a MISO or SIMO channel, layered broadcast coding achieves:
\begin{align}
\Delta \triangleq - \lim_{P\rightarrow\infty} \frac{\log \EH[D]}{\log P}
       = \min(b,L),
\end{align}
where $L$ is the total diversity order from independent fading blocks and antennas.
Moreover, the layered broadcast coding distortion exponent is shown to be optimal and CSIT does not improve $\Delta$, whereas diversity increases $\Delta$ up to a maximum as limited by the bandwidth ratio $b$. 

%%% ============================================================
\section{Conclusion}
\label{sec:conclu}

We considered the problem of source-channel coding over a delay-limited fading channel without CSI at the transmitter, and derived the optimal power distribution that minimizes the end-to-end expected distortion in the layered broadcast coding transmission scheme with successive refinement.
In the case when the channel undergoes Rayleigh fading with diversity order $L$, the optimal power distribution is congregated around the middle layers, and within this range the lower layers are assigned more power than the higher ones. As SNR increases, the power distribution of the higher layers remains unchanged, and the extra power is allocated to the idle lower layers.
Furthermore, increasing the diversity $L$ concentrates the power distribution towards the expected channel power gain $\bar{\gamma}$, while a larger bandwidth ratio $b$ spreads the power distribution further into the lower layers.
On the other hand, in the limit as $b$ tends to zero, the optimal power distribution that minimizes expected distortion converges to the power distribution that maximizes expected capacity.
While the expected distortion can be improved by acquiring CSIT or increasing the diversity order, it is shown that at high SNR the performance benefit from diversity exceeds that from CSIT, especially when the bandwidth ratio $b$ is large.

%%% ============================================================
\bibliographystyle{ieeebib/IEEEtran.bst}
\bibliography{ieeebib/IEEEabrv,bib/wrlscomm}

\end{document}